# Soft Modes and Relaxor Ferroelectrics

R. A. Cowley[1], S. N. Gvasaliya[2,*] and B. Roessli[2]

[1] Clarendon Laboratory, University of Oxford, Parks Road , Oxford OX1 3PU
[2] Laboratory for Neutron Scattering ETH Zurich and Paul Scherrer Institut, CH-5232, Villigen PSI, Switzerland

**Abstract**

Relaxor ferroelectrics are difficult to study and understand. The experiment shows that at low energy scattering there is an acoustic mode, an optic mode, dynamic quasi-elastic scattering and strictly elastic scattering as well as Bragg peaks at the zone centre. We have studied the scattering using the TASP spectrometer at PSI and have analysed the data using a model with interactions between the different components particularly to determine the properties of the elastic scattering. The quasi-elastic scattering begins to become significant at the Burns temperature of 620 K. It steadily increases in intensity on cooling reaching a maximum at ~400 K. Below this temperature the strictly elastic scattering begins to increase and shows a broadened line shape characteristic of crystals in a random applied field. We show that all the results obtained from PMN for the elastic scattering are consistent with the crystal having a random field transition at ~400 K. We have obtained similar results for PMN-PT and PZN-PT suggesting that random fields of the nano-regions also play an important role in these materials.

* On leave from Ioffe Physical Technical Institute, 26 Politekhnicheskaya, 194021 St. Petersburg, Russia

## 1. Introduction

The physical properties of the relaxor ferroelectrics have been intensively studied over the past 20 years. The dielectric constant has a broad maximum and there is not necessarily a well defined transition followed by a ferroelectric phase at lower temperature to a ferroelectric state. There have been many studies with x-ray and neutron scattering techniques as well as other techniques to determine the origin of these unusual properties. However, as occurred also with the high temperature superconductors, the lack of even a qualitative theoretical picture has resulted in a large quantity of experimental data and it is now difficult to know which results are important and which are not.

Since these materials are ferroelectric and have structures similar to those of many other regular ferroelectrics many of the initial experiments focused on trying to identify the soft mode for the phase transition in analogy to experiments on $SrTiO_3$, for example. Measurements were made of the phonon spectra at energy transfers less than but comparable with the temperature of the transition, namely 400-600 K. This approach discovered many interesting results such as the waterfall effect but were unsuccessful in identifying the soft mode. More recently several experimental groups have used cold neutrons and much better energy resolution than could be obtained with thermal neutron spectrometers. They have found that the soft mode is in the regions of 0.5 meV, two orders of magnitude below the initially expected energy for the soft mode. Furthermore it has been found that there is strong elastic scattering that has a varying wave vector dependence with temperature and so consists of several different components. A further complication is that the samples can show history dependent behaviour so that the results obtained at a particular temperature and applied electric field depend on how these parameters were approached.

We consider that most of these results are consistent with the behaviour to be expected from a system with random applied fields. Since systems with random applied fields have mostly been studied in magnetism the next section gives a brief review of the results obtained for magnetic systems and we shall describe how the response occurs at low energies, how random fields destroy long range order and how they can give rise to history dependent effects. In the third section we shall concentrate on the experimental results obtained for PMN, $PbMg_{1/3}Nb_{2/3}O_3$, and then in the fourth section on the materials such as PMN doped with PT, $PbTiO_3$, and PZN, $PbZn_{1/3}Nb_{2/3}O_3$ doped with PT. In the final section we shall summarise our results and how they can be explained as arising from random fields.

## 2. Random Fields in Magnetism

In magnetism random fields are not usually produced when defects are present in magnetic material because the lack of magnetic monopoles makes the formation of random local fields difficult. However, there is a very useful trick pointed out by Aharony and Feldman [1] that when a uniform magnetic field is applied to a random-uniaxial Ising antiferromagnet, the field induces a random-uniaxial staggered field. Because the applied field can be varied in strength this has enabled measurements to investigate the properties as the strength of the random fields was controlled and varied. The effect of random fields on the ordering and phase transitions of Ising systems was investigated theoretically by Imry and Ma [2] who showed that in two-dimensional systems the random fields always destroyed the long range order while in three-dimensions the phase transition changed to having different exponents but there was still long range order at low temperatures. The experiments confirmed the predictions for two-dimensions [3], but in three dimensions [4] the experiments showed there was a line in the temperature-field space below which the system was frozen and above which it was ergodic and fluctuating. A field cooled system did not then exhibit long range order but a zero field cooled system always did. Since this was different from the theoretical predictions there followed a 10 year period of discussion between the theoreticians and the experimentalists. It was settled by experiments with x-ray and neutron scattering techniques showing that zero field cooled experiments [5] had extremely long relaxation times and that it was not possible to observe the true phase transition. It was concluded that the failure to observe long range order in the bulk was because of the difficulty the sample had in developing long range order even though it had the lowest free energy. These results showed how random fields produced transitions with unusual properties, domains that were very slowly relaxing and results that were history dependent.

Imry and Ma [2] also developed the theory to describe continuous symmetry materials such as Heisenberg models. They showed that if the random fields had continuous symmetry the random fields would destroy the long range order at any temperature or strength of field for a three dimensional material. For these systems the trick of Aharony and Feldman [1] cannot be used because a uniform field breaks the continuous symmetry. Consequently in magnetism continuous symmetry random fields cannot be easily studied experimentally. Relaxor ferroelectrics are cubic materials and the random positions and different charges of the Mg and Nb ions produce electric fields whereas in magnetism defects only produce dipoles. Since the symmetry of relaxors is close to continuous these random fields may be able to destroy the long range order. On the other hand cubic symmetry is not continuous and so ordering may be possible when the cubic anisotropy is sufficiently large. These are the ingredients that we and others have used to suggest a random field theory for relaxors [6,7,8] although a complete theory for relaxor ferroelectrics must include the low energy soft mode, several phase transitions, nearly long range order and history and surface dependent effects.

## 3. The Experimental Results for PMN

Experiments to study the dynamics and statics of the relaxor ferroelectrics are difficult and there has been only slow progress in developing an understanding. The difficulty is because they are ferroelectric and so the soft mode is at the zone centre and the scattering from this must be distinguished from the Bragg peak and the acoustic phonons. In fact, it is even more complex because in the relaxors, a crucial role is also played by the transverse optic mode and the elastic diffuse scattering. Initially the experiments began with a study of the scattering from the acoustic and lowest energy optic and acoustic modes. This was because the typical phase transition temperatures as shown by the dielectric susceptibility are about

400 K and consequently this suggested that the energy of the soft mode might be around 40 meV [9]. The experiments then shifted to study the elastic diffuse scattering and it was found that it was peaked at the Bragg peaks but its shape in wave vector varied with temperature and also changed from one Brillouin zone to another [10]. Both of these studies produced very interesting results but we do not here have the space to discuss the interesting results from either of these sets of experiments. The most recent development has been to study the dynamics of the elastic scattering and the results have been crucial for developing an understanding of the mechanism of the phase transition in relaxors. There are two ways in which the experiments can be done. We [8] have used cold neutron scattering, $E_f = 5$ meV and then analysed the spectra to determine the properties by using an interacting model consisting of transverse optic mode, acoustic mode, dynamic quasi-elastic mode and strictly elastic scattering. This complex model has a large number of parameters but, with the aid of symmetry, the systematic behaviour can be deduced from the results and typical results are shown for the scattering from PMN in fig. 1. The alternative approach is to use novel neutron scattering techniques such as back-scattering [11] and spin-echo spectrometers to obtain the resolution needed to study the nature of the elastic scattering. The results of both of these techniques agree with one another at least qualitatively.

Figure 1 shows the scattering obtained from PMN for a reduced wave-vector of (1,1,0.05) close to the Bragg reflection at (1,1,0) at three different temperatures, 670 K above the Burns temperature of 620 K, 430 K well below the Burns temperature and 300 K is an even lower temperature. The solid lines are fits to the spectra and the dotted lines are the scattering from the fits that cannot be described by resolution-limited elastic scattering and result from the dynamic quasi-elastic scattering. The spectrum at a temperature 670 K has a small strictly elastic component and no quasi-elastic scattering. The strictly elastic scattering arises from the incoherent scattering and from the site disorder from the random positions of the Mg and Nb ions. At a temperature of 430 K the strictly elastic scattering has increased by a factor of less then 2 while there is a large and distinct quasi-elastic component. At lower temperatures the quasi-elastic scattering is about the same intensity but the strictly elastic scattering has increased by about a factor of 20.

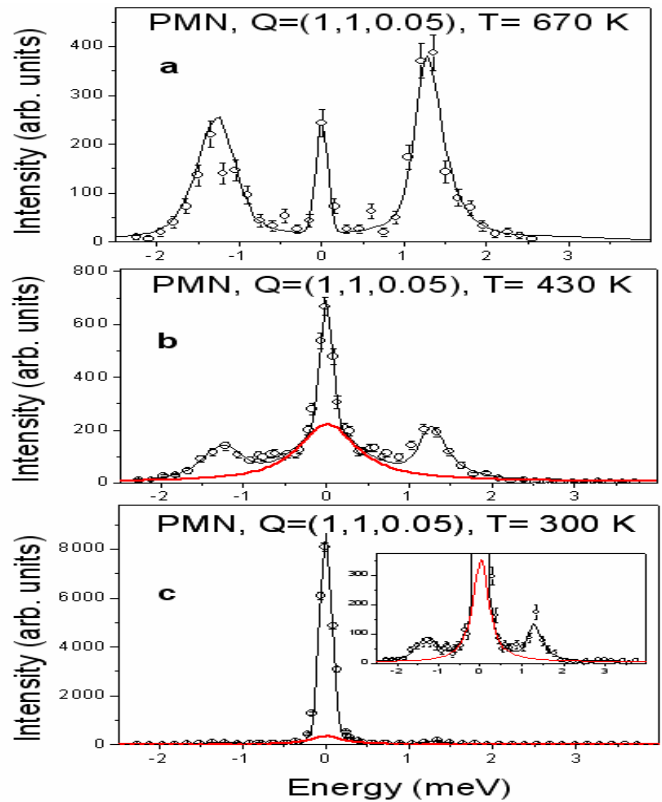

Figure 1: The constant Q scattering from PMN at 3 temperatures. The solid line is the fit to the model and the dotted line is the quasi-elastic component

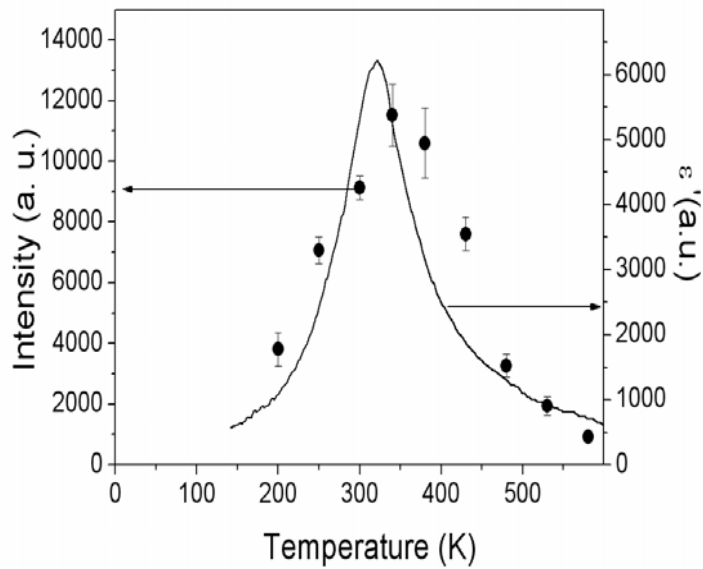

Figure 2 The intensity of the quasi-elastic scattering at a wave vector of (1,1,0.05) and the dielectric susceptibility.

These measurements were made over a range of temperatures and wave vectors and the intensity of the elastic and quasi-elastic scattering for a wave vector transfer of (1,1,0.05) is shown in fig. 2. The quasi-elastic scattering has a peak at a temperature of about 350 K which is approximately the temperature of the maximum dielectric response while the elastic scattering is small above a temperature of 400 K and then increases on further decreasing the temperature. This behaviour suggests that there is a phase transition at about 400K and that above this temperature PMN has only dynamic ferroelectric fluctuations while below 400K there are both dynamic fluctuations and a static or very slow ferroelectric ordered component. The Burns temperature is about 620 K much higher than this ordering temperature but is consistent with the temperature at which the quasi-elastic scattering begins to become significant. We suggest that this appearance of quasi-elastic scattering is associated with the onset of dynamic polar regions and that the scattering from these steadily increases on cooling.

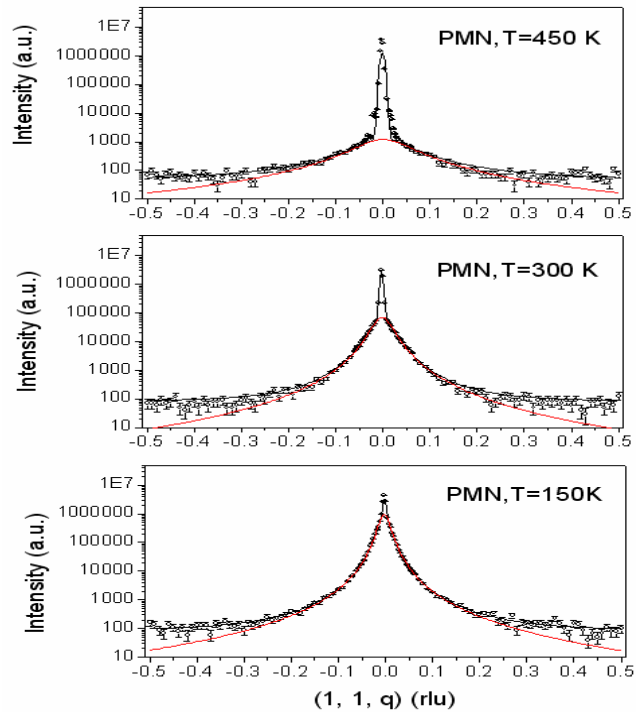

Figure 3 The scattering through the (1,1,0) Bragg peak at 3 temperatures.

The phase transition is not a conventional structural phase transition because there is no onset of long range ferroelectric order as shown by the elastic scattering in fig.3. Below the phase transition temperature of 400 K there is an intense elastic component that is increasing in intensity on cooling and which is strongly peaked at the Bragg position. The shape of this peak is well described not by a simple Lorentzian or Gaussian but by a Lorentzian to the power of 1.5 and the width of the Lorentzian decreases from 0.025 rlu at a temperature of 300 K to 0.015 rlu at 150 K The behaviour of the dynamic quasi-elastic scattering has been further investigated and the results are shown in fig. 4 for the correlation length and fig 5 for the dynamical width. The correlation width of the scattering decreases on
cooling and becomes almost constant below the transition temperature whereas the dynamical width also decreases with decreasing temperature and in this case continues to decrease below the transition temperature. These features are fully consistent with a random field transition occurring at about 400 K. The order parameter does not have long range order but a distinctive shape peaked strongly at the Bragg peak wave vector with a shape given by a Lorentzian to a larger than usual power but the same power as observed at magnetic transitions in random fields while the correlation length for the dynamic scattering is not infinite at the transition temperature as observed at magnetic transitions. There is also evidence that the surface behaviour of PMN is different from the bulk as found at magnetic materials in random fields [12]

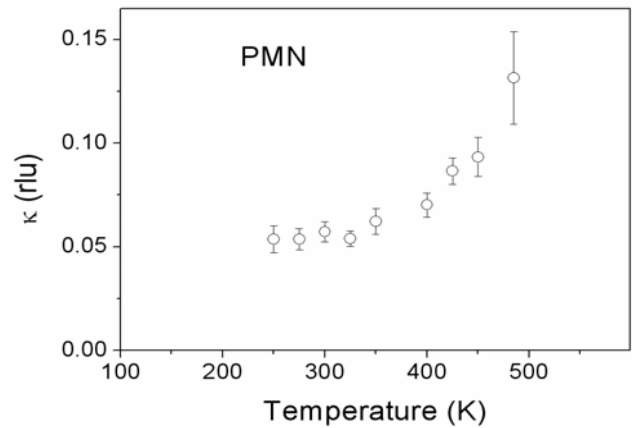

Figure 4  Inverse correlation length for the quasi-elastic scattering from PMN.

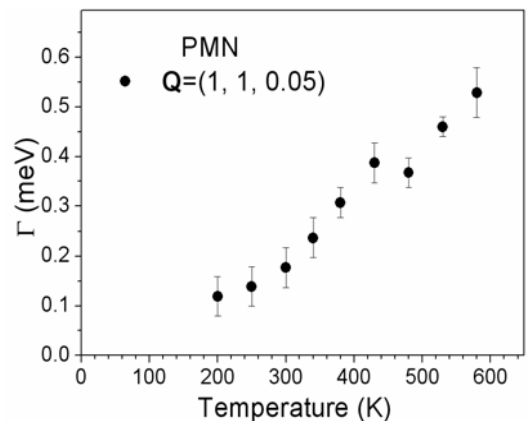

Figure 5 Damping constant of the quasi-elastic scattering for PMN at (1,1,0.05)

Below a temperature of 200K PMN can be poled if an electric field is applied and there have been many experiments to study the structure [13]. We consider this is possibly similar to the uniaxial random field behaviour and suggest that at this temperature the structure changes from a uniform isotropic state which cannot support long range order to a state where the cubic anisotropy of the system is sufficiently large that the ground state becomes a long range ferroelectric state. This is not achieved at very low electric fields because of the difficulty of achieving thermal equilibrium in random field states.

### 4. Relaxors and Ferroelectricity

PMN and PZN doped with sufficient PT become ferroelectric when cooled and the transition is accompanied by a large piezoelectric constant that makes the materials of technical interest. Experiments have been performed [14] to try to determine the nature of the ferroelectric transition and whether it is similar to that of relaxors such as PMN or that of pure ferroelectrics. We have studied a sample of 68%PMN-32%PT with an expected transition temperature of about 370 K From the point of view in this article the important question is the nature of the elastic scattering and this is shown in fig.6.

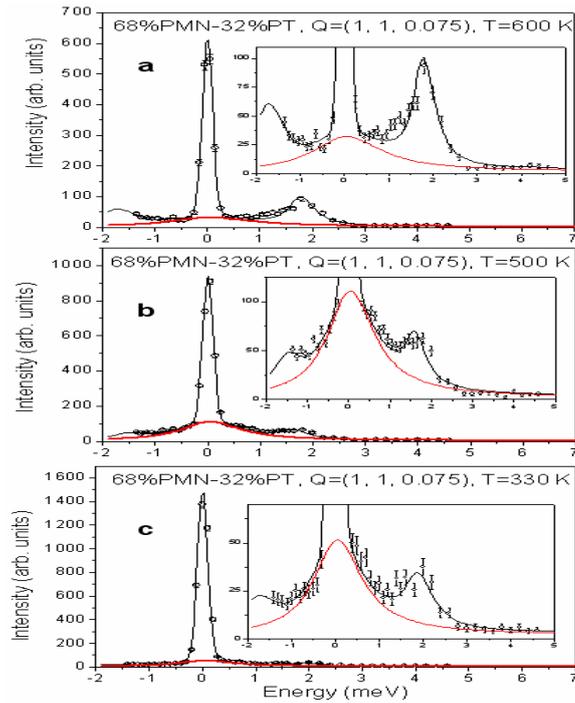

Figure 6 Scattering from PMN-PT showing the fits to the model. The solid line shows the fit and the dotted line the quasi-elastic component.

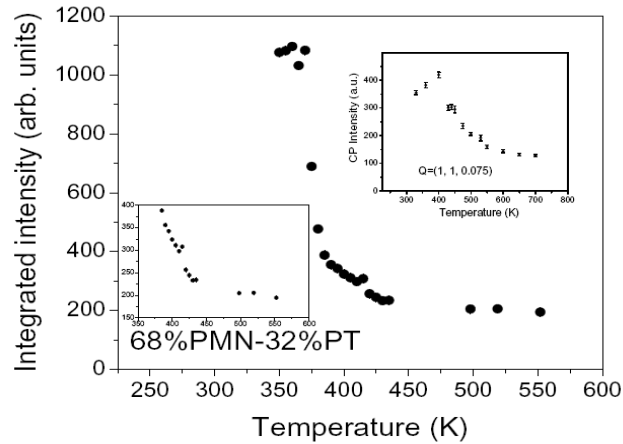

Figure 7 Temperature dependence of the strictly elastic scattering from PMN-PT showing the scattering at (0,0,2) and at (1,1,0.75)

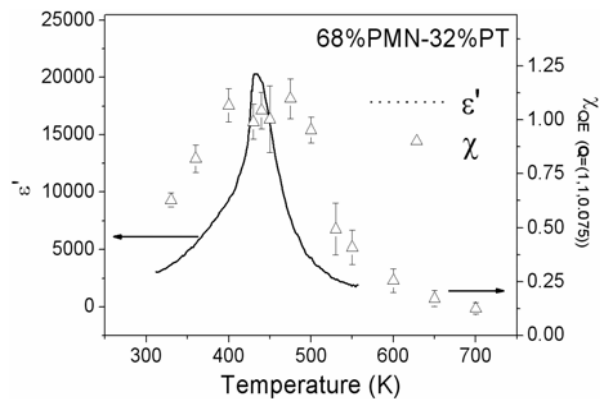

Figure 8 Intensity of the quasi-elastic scattering at Q=(1,1,0.05) and the dielectric susceptibility for PMN-PT

Both quasi-elastic and strictly elastic components are present at all three temperatures suggesting that the behaviour is very similar to that of PMN. In fig 7 the temperature dependence of the strictly elastic scattering is shown. This scattering away from the Bragg reflection has a maximum at a temperature of about 400 K whereas at the Bragg reflection there is a rapid change occurring about 370 K and the intensity then steadily increases with decreasing temperature. The quasi-elastic scattering away from a Bragg peak is shown in fig 8 and has a peak centred on a temperature of 450K which is very similar to the peak of the dielectric susceptibility. These measurements are consistent with the random phase transition occurring at a temperature of about 450 K while the ferroelectric transition occurs at a temperature of 370 K because the symmetry breaking ferroelectric fluctuations are much stronger than in PMN. Measurements have been made of the inverse correlation length and damping of the quasi-elastic scattering in this material and the results are qualitatively similar to those described above for PMN although differing in detail. Similar results [15] showing both quasi-elastic scattering and strictly elastic scattering have been observed in 93%PZN-7%PT. These results are different [16] from those of other authors who did not observe any quasi-elastic scattering in these materials but their measurements had less resolution because they used a larger incident neutron energy.

## 5. Conclusions

The understanding of relaxor ferroelectrics is difficult, complex and uncertain. We have argued that at least a qualitative understanding of the behaviour can be had from a detailed study of the quasi-elastic and strictly elastic scattering. This understanding does not state that the understanding of the phonons is unimportant. They undoubtedly influence the properties because the measurements show that the optic mode decreases in energy towards the phase transition and that the transverse optic mode becomes more highly damped in the region of the phase transition [9]. Likewise the shape of the diffuse elastic scattering provides information about which atoms are contributing to the polar nano-regions. These are both important aspects of these materials and require more study. Our point of view is that the essential understanding must however come from the quasi-elastic scattering and the associated strictly elastic scattering that are difficult to separate and measure. We alone have achieved this by the use of cold neutron scattering, when a quantitative analysis of the measurements has shown for PMN that the Burns temperature is connected with the onset of quasi-elastic scattering. Above this temperature there is little strictly elastic scattering from the different scattering lengths of Mg and Nb ions but below the atoms displace away from the nominal lattice sites and increasingly form polar nano-regions giving rise to dynamic quasi elastic scattering. On cooling to ~ 420 K the system then undergoes a random field transition and there is an increasing amount of static short range order while the dynamic scattering has a maximum in intensity and the correlation length becomes constant below this temperature. We suggest that the behaviour below 200K can be understood if the stable state changes from one with isotropic short range order to one with cubic anisotropy which can support a ferroelectric state if the material is polarised by a symmetry breaking electric field. Similar results with both quasi-elastic and strictly elastic scattering varying with temperature have been found in PMN doped with PT and PZN doped with PT.

These experiments are difficult because it is impossible to measure the elastic scattering close to a Bragg reflection. Consequently some of our results are measurements taken close to the Bragg reflection when the quantity of theoretical interest is the behaviour at the Bragg reflection. This problem causes considerable experimental difficulty and we innovative techniques are required to extract the behaviour and compare with theoretical models. We look forward to the development of new ways of obtaining the necessary data.

**Acknowledgements**

We are grateful to G. M. Rotaru, S. Kojima, S. G. Lushnikov , P. Gunter, P. Huber for assistance with the experiments and to them and C. Stock for helpful discussions. Financial support was provided by the Leverhulme foundation. This work is based on experiments performed at the Swiss spallation neutron source SINQ, Paul Scherrer Institute, Villigen, Switzerland.